\documentclass{article}
\usepackage{graphicx} 
\usepackage{amsmath}
\usepackage{amssymb}
\usepackage[numbers]{natbib}
\usepackage{appendix}
\usepackage{authblk}

\title{Lagrange's planetary equations with time-dependent secular perturbations}
\author[1]{Barnab{\'a}s Deme}
\affil[1]{CNRS and Sorbonne Universit{\'e}, UMR 7095, Institut d'Astrophysique de Paris, 98 bis Boulevard Arago, F-75014 Paris, France}

\begin{document}

\maketitle

\begin{abstract}
    The long-term evolution of astrophysical systems is driven by a Hamiltonian that is independent of the fast angle. As this Hamiltonian may contain explicitly time-dependent parameters, the conservation of mechanical energy is not guaranteed in such systems. We derive how the semi-major axis evolves in these cases. We analyze two astrophysically interesting examples, those of the harmonic and quadrupole perturbations. 
\end{abstract}

\section{Introduction}
Lagrange's planetary equations describe the time evolution of the orbital elements of a Keplerian ellipse: the semi-major axis ($a$), the eccentricity ($e$), the inclination ($i$), the argument of the periapsis ($\omega$), the length of the ascending node ($\Omega$) and the epoch of the periapsis passage ($t_0$). Although they suffer from singularities at some orbital configurations ($i=0$, $e=0$, $e=1$),\footnote{These singularities arise from the coordinate-dependence of the equations, i.e., we have to set a frame of reference in which we define the orbital elements. They can be overcome by using vectors, e.g. the angular momentum and eccentricity vectors \citep{milankovitch39,klein1924}.}  they have been efficiently used in celestial mechanics for many different problems \citep[e.g.,][]{murraydermott99,tremaine}.

These equations can be derived in different ways. For example, Ref.~\citep{brouwer1961} calculates Lagrange brackets to obtain them. Alternatively, Ref.~\citep{valtonen06} gets these equations from the time derivatives of Delaunay's canonical variables (see Appendix \ref{appendix} for a short summary). Normally it is assumed that the interacting bodies' masses are constant during the evolution. In this Letter, we omit this assumption, which is motivated by astrophysical processes (e.g., losing or gaining mass via winds or accretion \citep{illes2024,mustill2024}). More importantly, we show that \textit{perturbations in general can be modeled as (potentially time-dependent) changes in the masses.} A similar idea is used in the case of radiation pressure \citep{burns1979}, but here we treat a much broader class of perturbations,\footnote{The concept that perturbations effectively change parameters as mass or charge, is frequently used in plasma and stellar dynamics \citep{hamilton2024} and in quantum field theory \citep{gifted}.} e.g., harmonic and quadrupole ones.

In many cases one is not interested in the evolution of the orbital elements on the timescale of the orbital period. Instead, one wishes to know the change of the elements that is accumulated over many orbital cycles. This is called the secular approximation \citep[e.g.,][]{naoz2016}. The evolution is so slow (adiabatic) that the orbit can be considered unperturbed during one period. It means that the initial orbital phase has no impact on the evolution, i.e., one can average the equations over the fast angle (mean anomaly) \citep{vonzeipel10}.

In this Letter we investigate secular dynamics in the case where the perturbation potential has an adibatically evolving parameter. In Sec.~\ref{sec:lagrange} we rewrite Lagrange's planetary equations so that they account for explicitly time-dependent masses. We assume that this time-dependence is adiabatic, so it allows us to execute orbit-averaging. In Sec.~\ref{sec:effective} we model perturbations as a shift in the mass parameter. In Sec.~\ref{sec:examples} we derive and numerically solve Lagrange's planetary equations with secular time-dependent perturbations. In Sec.~\ref{sec:discussions} we conclude.

\section{Lagrange's planetary equations}\label{sec:lagrange}
Appendix~\ref{appendix} introduces the notations and  contains the derivation of Lagrange's planetary equations. We focus on the secular evolution, so we drop terms that depend on the fast angle ($\partial \mathcal{H}/\partial M=0$). We investigate central force field problems only,\footnote{As central problems are always planar, we restrict ourselves to 2D motion only. For the possibility of generalizations to 3D, see Sec.~\ref{sec:discussions}.} which are axisymmetric ($\partial \mathcal{H}/\partial \omega=0$). With these restrictions, Eqs.~\eqref{eq:dot_a}-\eqref{eq:dot_omega} become
\begin{equation}\label{eq:dot_a_short}
    \dot{a}=-\frac{a}{\mu}\dot{\mu},
\end{equation}
\begin{equation}\label{eq:dot_e_short}
    \dot{e}=0,
\end{equation}
and
\begin{equation}\label{eq:dot_omega_short}
    \dot{\omega} = -\frac{\sqrt{1-e^2}}{e\sqrt{\mu a}}\frac{\partial \mathcal{H}}{\partial e}.
\end{equation}
Note that the term on the right-hand side of Eq.~\eqref{eq:dot_a_short} is usually not accounted for (e.g., Eq. (6.145) in Ref.~\citep{murraydermott99}). We immediately see that the actions $L$ and $G$ in Eqs.~\eqref{eq:Ham_lL} and \eqref{eq:Ham_gG} are first integrals of these equations which is in agreement with our expectation that the actions are adiabatic invariants \citep{goldstein02,lichtenberg,arnold78}. It means they are conserved if the parameter $\mu$ changes adiabatically, i.e. on secular timescales: $\dot{\mu}/\mu\ll \mu^{1/2}a^{-3/2}$. 

\section{Perturbation as effective mass}\label{sec:effective}
The Hamiltonian in Eq.~\eqref{eq:dot_omega_short} consists of an unperturbed and a perturbing part:
\begin{equation}
    \mathcal{H}=\frac{\dot{x}^2+\dot{y}^2}{2}-\frac{\mu}{r}+\epsilon(t) V(r),
\end{equation}
where $r=\sqrt{x^2+y^2}$, $\epsilon(t) \ll 1$ is a small parameter\footnote{In fact, we demand it to be small initially: $\epsilon(0)\ll 1$. In Sec.~\ref{sec:discussions} and Appendix~\ref{sec:app_smallness} we show that it can approach unity at later times without violating the perturbative assumption.} and $V(r)$ is a central potential field perturbation. The equations of motion are
\begin{equation}
    \ddot{x}=-\frac{\mu x}{r^3}-V'\frac{x}{r}
\end{equation}
and similarly for $y$, where the ' denotes differentiation with respect to the argument. We rewrite this equation as
\begin{equation}
    \ddot{x}=-\frac{ \left( \mu+V'r^2 \right)x}{r^3}
\end{equation}
and similarly for $y$. As the term $\mu+V'r^2$ appears formally as a new mass parameter in these equations, we model the perturbations as effective changes in the mass parameter $\mu$. More precisely, since we are interested only in the secular behavior, we define the effective mass as the orbit-averaged value, i.e.
\begin{equation}
    \mu_\mathrm{eff}=\langle \mu+V'r^2 \rangle =\frac{1}{2\pi}\int_0^{2\pi} \left( \mu+V'r^2\right) \mbox{ d}M.
\end{equation}
Equations~\eqref{eq:dot_a_short} and \eqref{eq:dot_omega_short} can now be rewritten as\footnote{We ignore the equation of the eccentricity, because it does not evolve in our approximation.}
\begin{equation}\label{eq:dot_a_final}
    \dot{a}=-\frac{a}{\mu_\mathrm{eff}}\dot{\mu}_\mathrm{eff},
\end{equation}
and
\begin{equation}\label{eq:dot_omega_final}
    \dot{\omega} = -\frac{\sqrt{1-e^2}}{e\sqrt{\mu a}}\frac{\partial \langle \mathcal{H}\rangle}{\partial e},
\end{equation}
which drive the secular evolution of the orbital elements in the presence of any central potential field perturbations. Note that $\mu$ is not to be changed to $\mu_\mathrm{eff}$ in the equation of $\dot{\omega}$.

\section{Time-dependent secular perturbations}\label{sec:examples}

\subsection{Time-dependent harmonic perturbations}\label{sec:example_harmonic}
Now we test the validity of Eqs.~\eqref{eq:dot_a_final}-\eqref{eq:dot_omega_final}. As a first example, we consider the following Hamiltonian:
\begin{equation}\label{eq:quadrupole_hamiltonian}
    \mathcal{H}=\frac{\dot{x}^2+\dot{y}^2}{2}-\frac{\mu}{r}+\epsilon(t) k_1r^2,
\end{equation}
where $k_1$ is a constant of dimension $\mathrm{energy}/\mathrm{distance}^2$. It can model e.g. orbits around a supermassive black hole embedded in a dark matter halo core \citep{kormendy,deblok}, or around an intermediate-mass black in the core of a globular cluster \citep{haberle2024,noyola2006a}.\footnote{Although the presence of an intermediate-mass black hole may turn the core to a cusp. \citep{noyola2006b}} Both environments may be subject to time-evolution. 

The effective mass is
\begin{equation}
    \mu_\mathrm{eff}=\frac{1}{2\pi}\int_0^{2\pi} \left(\mu+2\epsilon k_1r^3\right) \mbox{ d}M =  \mu+2\epsilon k_1 a^3f(e),
\end{equation}
where $f(e)=\left( 1+3e^2+\frac{3}{8}e^4 \right)$
We assume that the strength of the perturbation changes adiabatically: $\dot{\epsilon}/\epsilon\ll \mu^{1/2}a^{-3/2}$. Plugging this into Eqs.~\eqref{eq:dot_a_final}-\eqref{eq:dot_omega_final} yields
\begin{equation}\label{eq:adot_harmonic}
    \dot{a}=-\frac{2\dot{\epsilon}k_1f(e)a^4}{\mu+8\epsilon k_1 a^3 f(e)}
\end{equation}
and
\begin{equation}\label{eq:omegadot_harmonic}
    \dot{\omega}=-3\epsilon k_1 \sqrt{\frac{a^3(1-e^2)}{\mu}}.
\end{equation}

\begin{figure*}
\centering
\includegraphics[scale=0.6]{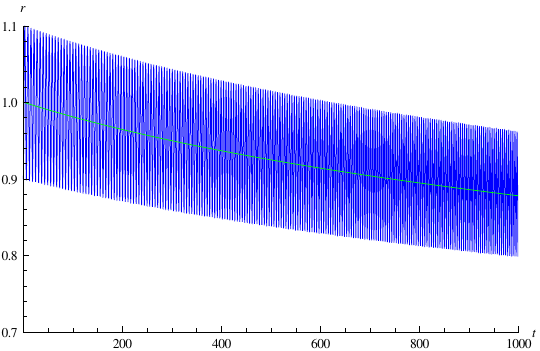}
\includegraphics[scale=0.6]{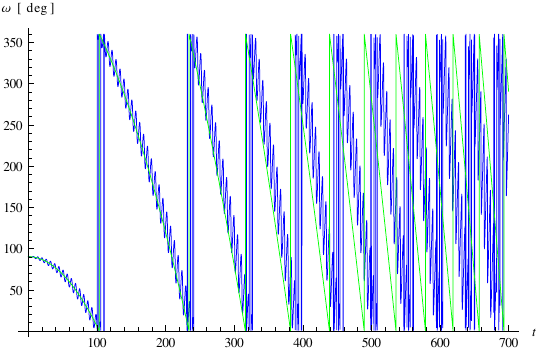}
\caption{Numerical solution of the direct equations of motion (blue) and of Eqs.~~\eqref{eq:adot_harmonic}-\eqref{eq:omegadot_harmonic} (green). Units are set by the choice $\mu=k_1=1$ (note that time and distance are dimensionless in this system of units). The perturbation strength evolves as $\epsilon=10^{-4}t$. The initial conditions are $a_0=1$, $\omega_0=90^\circ$ and $e_0=0.1$. \textit{Left:} The distance $r$ from the center, the mean of which is the semi-major axis. \textit{Right:} The argument of pericenter $\omega$. It precesses quickly at the end of the simulation which results in a phase shift from the curve predicted by the theory. Note that it is truncated at $t=700$ for better visibility.}
\label{fig:harmonic}
\end{figure*}

Figure~\ref{fig:harmonic} shows the solution of Eqs.~\eqref{eq:adot_harmonic}-\eqref{eq:omegadot_harmonic} along with the result of the direct numerical integration of the equations of motion. The agreement between the curves is good. The rapid oscillation of the blue curve originates from the orbital motion, which is averaged out from the secular green curve. Note that on the vertical axis of the left plot we show the distance $r$ from the center. Its mean is the semi-major axis $a$, while its width is $2ae$. Since the eccentricity is constant (Eq.~\ref{eq:dot_e_short}) and the relative variation of $a$ is moderate ($\Delta a/a\sim 0.1$), the amplitude of the $r$-oscillation remains roughly constant. 

\subsection{Time-dependent quadrupole}
Our second example is the following Hamiltonian:
\begin{equation}
    \mathcal{H}=\frac{\dot{x}^2+\dot{y}^2}{2}-\frac{\mu}{r}+\epsilon (t) k_2 r^{-3},
\end{equation}
where $k_2$ is a constant of dimensions $\mathrm{energy}\times\mathrm{distance}^3$. It may model the motion around an oblate body in its equatorial plane \citep{faridani2023,stepien2002,beletsky} or around a tight inner binary in its orbital plane \citep{standing2023}.

The effective mass is 
\begin{equation}
    \mu_\mathrm{eff}=\frac{1}{2\pi}\int_0^{2\pi} \left(\mu-3\epsilon k_2r^{-2} \right)\mbox{ d}M =  \mu -\frac{3\epsilon k_2}{a^2\sqrt{1-e^2}}.
\end{equation}
It results in 
\begin{equation}\label{eq:adot_quadrupole}
    \dot{a}=\frac{3\dot{\epsilon}k_2}{\mu a\sqrt{1-e^2}+3\epsilon k_2/a}
\end{equation}
and
\begin{equation}\label{eq:omegadot_quadrupole}
    \dot{\omega}=-\frac{3\epsilon k_2}{\mu^{1/2}a^{7/2}(1-e^2)^2}.
\end{equation}

\begin{figure*}
\centering
\includegraphics[scale=0.6]{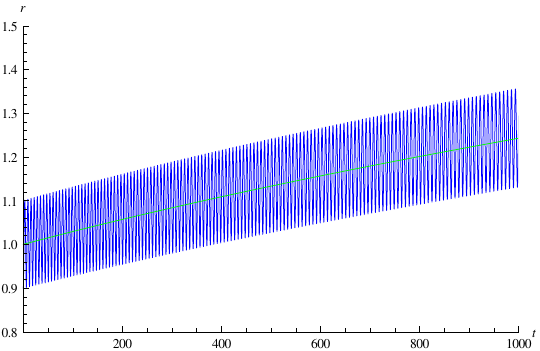}
\includegraphics[scale=0.6]{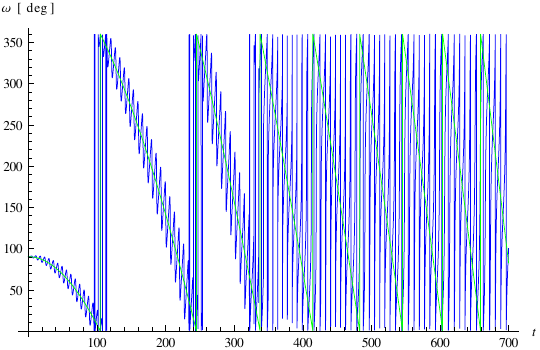}
\caption{Same as Fig.~\ref{fig:harmonic}, but for the quadrupole perturbation and with $\mu=k_2=1$ (time and distance units are set to 1).}
\label{fig:quadrupole}
\end{figure*}

Similarly to the case with the harmonic perturbation, Fig.~\ref{fig:quadrupole} shows the solution of Eqs.~\eqref{eq:adot_quadrupole}-\eqref{eq:omegadot_quadrupole}. It is again in a good agreement with the result of the direct integration. 

\section{Discussion}\label{sec:discussions}
\begin{figure*}
\centering
\includegraphics[scale=0.6]{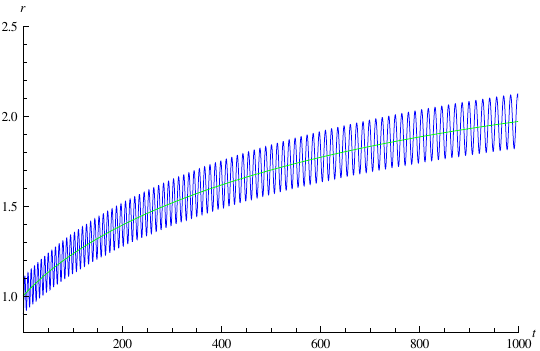}
\caption{Same as the left panel of Fig.~\ref{fig:quadrupole}, but with a perturbation evolving as $\epsilon=1-\mathrm{e}^{-0.001 t}$. The amplitude of the oscillations grows because of the significant change in the semi-major axis.}
\label{fig:quadrupole_exp}
\end{figure*}

Actions are well-known as adiabatic invariants: they remain constant if the physical parameter in the Hamiltonian (e.g. the mass parameter $\mu$) changes slowly. Here we showed how the \textit{unperturbed} Hamiltonian's action (or more precisely, the semi-major axis) evolves when the parameter in the \textit{perturbation} (e.g. the coupling strength $\epsilon$) varies in time. The validity of the theory, similarly to that of the standard adiabatic invariants, does not depend on how the parameter evolves in time. The only requirement is that it must be adiabatic. Indeed, Figs.~\ref{fig:harmonic}
-\ref{fig:quadrupole} show linear, while Fig.~\ref{fig:quadrupole_exp} shows exponential time evolution.

Secular problems, i.e. those that occur on long timescales compared to the orbital period, are usually treated via von Zeipel's transformation \citep{vonzeipel10}. In Appendix~\ref{sec:app_vonzeipel} we show that in systems with secular time-dependent perturbations von Zeipel's transformation is inconvenient compared to the method we suggest. Its main disadvantage is that one needs to perform several consecutive transformations to increase the accuracy, while the method described above does not need it.\footnote{The main reason of this is that the actions can be adiabatic invariants up to high orders in the "slowness" parameter $\dot{\epsilon}/(\mu^{1/2}a^{-3/2})$. For a similar discussion about the harmonic oscillator, see Ref.~\citep{kulsrud1957}.}  

\begin{figure*}
\centering
\includegraphics[scale=0.6]{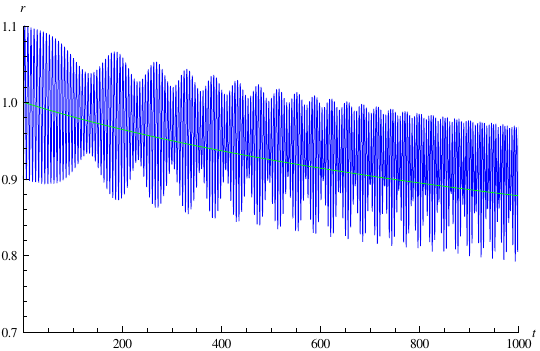}
\caption{Same as the left panel of Fig.~\ref{fig:harmonic}, but with anisotropic harmonic perturbation.}
\label{fig:quadrupole_ani}
\end{figure*}

The biggest caveat of our work is that it works only for central potential perturbations, i.e. those that depend only on $r$. The reason for this is that in anisotropic cases each direction may have its own $\mu_\mathrm{eff}$ and it is not obvious which of them one should use in Eq.~\eqref{eq:dot_a_final}. However, Eqs.~\eqref{eq:dot_a_final}-\eqref{eq:dot_omega_final} may remain qualitatively correct even if the perturbation is anisotropic, as demonstrated in Fig.~\ref{fig:quadrupole_ani}. The blue curve in that figure is the solution of the (non-conservative) system
\begin{equation}
    \ddot{x}=-\frac{\mu x}{r^3}-2\epsilon k_1 x
\end{equation}
and
\begin{equation}
    \ddot{y}=-\frac{\mu y}{r^3}-\epsilon k_1 y.
\end{equation}
The modulation of the amplitude of the blue curve is the consequence of the angular momentum not being conserved anymore.

Another important point is the strength of the perturbation, $\epsilon (t)$. It must be small initially ($|\epsilon(0)| \lesssim 0.01$) for the orbit to be (quasi-) Keplerian. If it decreases, it hardly has any effect on the dynamics of the semi-major axis, because $|\Delta a| \sim |\Delta \epsilon| \sim 0.01$ (see, e.g., Eq.~\ref{eq:adot_harmonic}). However, if $|\epsilon|$ increases (which is not unphysical, e.g. dark matter halos gaining mass via accretion in the example of Sec.~\ref{sec:example_harmonic}), the change in the semi-major axis can be more significant: $|\Delta a| \sim |\Delta \epsilon| \gg 0.01$ (see Fig.~\ref{fig:quadrupole_exp}, where $\Delta a / a\sim 1$, which is also indicated by the $r$-amplitude becoming larger). Importantly, $\epsilon(t)$ itself does not need to remain small, only the ratio of the perturbing and unperturbed (Keplerian) energies. This latter statement holds, as proved heuristically in Appendix \ref{sec:app_smallness}.  

Finally, we emphasize that the machinery in Sec.~\ref{sec:effective} does not rely on the form of the unperturbed Hamiltonian. It can be applied to systems other than the Kepler problem.

\appendix
\section{Derivation of Lagrange's planetary equations}\label{appendix}
Here we follow the derivation in Ref.~\citep{valtonen06}. However, we constrain ourselves to planar motion.\footnote{Normally, $\omega$ is measured from the ascending node. In this 2D case, we arbitrarily define a reference direction.} We start with Delaunay's canonical variables: 
\begin{align}
    l=M, \quad  & L=\sqrt{\mu a},  \\
    g=\omega, \quad  & G=\sqrt{\mu a (1-e^2)},
\end{align}
where $\mu=\mathcal{G}m$ with $\mathcal{G}$ and $m$ being the gravitational constant and the mass of the one-center problem, respectively, and $M$ is the mean anomaly. These variables obey Hamilton's equations of motion:
\begin{align}
    \dot{l}=\frac{\partial \mathcal{H}}{\partial L}, \quad  & 
    \dot{L}=-\frac{\partial \mathcal{H}}{\partial L}, \label{eq:Ham_lL} 
    \\
    \dot{g}=\frac{\partial \mathcal{H}}{\partial G}, \quad  &
    \dot{G}=-\frac{\partial \mathcal{H}}{\partial g}, \label{eq:Ham_gG}
\end{align}
where the $\cdot$ denotes time derivative and $\mathcal{H}$ is the system's Hamiltonian. The defining relations above can be inverted to give
\begin{equation}\label{eq:inv_a}
    a=\frac{L^2}{\mu},
\end{equation}
\begin{equation}
    e=\sqrt{1-\frac{G^2}{L^2}},
\end{equation}
and
\begin{equation}\label{eq:inv_e}
    \omega=g.
\end{equation}
Lagrange's planetary equations are obtained by taking the time derivative of these expressions. In particular, using Hamilton's equations \eqref{eq:Ham_lL}-\eqref{eq:Ham_gG},
\begin{equation}\label{eq:dot_a_gen}
    \dot{a}=\frac{\partial a}{\partial \mu}\dot{\mu}+\frac{\partial a}{\partial L}\dot{L}=\frac{\partial a}{\partial \mu}\dot{\mu}-\frac{\partial a}{\partial L}\frac{\partial \mathcal{H}}{\partial l},
\end{equation}
\begin{equation}
    \dot{e}=\frac{\partial e}{\partial L}\dot{L}+\frac{\partial e}{\partial G}\dot{G}=-\frac{\partial e}{\partial L}\frac{\partial \mathcal{H}}{\partial l}-\frac{\partial e}{\partial G}\frac{\partial \mathcal{H}}{\partial g}
\end{equation}
and
\begin{equation}\label{eq:dot_omega_gen}
    \dot{\omega}=\dot{g}=\frac{\partial \mathcal{H}}{\partial G}.
\end{equation}
With the help of the conversion formulae \eqref{eq:inv_a}-\eqref{eq:inv_e}, we arrive at
\begin{equation}\label{eq:dot_a}
    \dot{a}=-\frac{a}{\mu}\dot{\mu}-2\sqrt{\frac{a}{\mu}}\frac{\partial \mathcal{H}}{\partial M},
\end{equation}
\begin{equation}
    \dot{e}=-\frac{1-e^2}{e\sqrt{\mu a}}\frac{\partial \mathcal{H}}{\partial M} + \frac{\sqrt{1-e^2}}{e\sqrt{\mu a}}\frac{\partial \mathcal{H}}{\partial \omega},
\end{equation}
and
\begin{equation}\label{eq:dot_omega}
    \dot{\omega} = -\frac{\sqrt{1-e^2}}{e\sqrt{\mu a}}\frac{\partial \mathcal{H}}{\partial e}.
\end{equation}

\section{Elimination of the harmonic perturbation by von Zeipel's method}\label{sec:app_vonzeipel}

We investigate the problem of Sec.~\ref{sec:example_harmonic} with von Zeipel's method. We rewrite Eq.~\eqref{eq:quadrupole_hamiltonian} as   
\begin{equation}
\mathcal{H}=-\frac{\mu^2}{2L^2} + J + \epsilon(\theta)k_1 r^2(l),
\end{equation}
where Delaunay's elements are used and we introduced an extra degree of freedom ($\theta$, $J$), such that the new angle plays the role of time: 
\begin{equation}
    \dot{\theta} = \frac{\partial \mathcal{H}}{\partial J}=1 \to \theta=\theta_0 + t.
\end{equation}
The generating function of von Zeipel's transformation is
\begin{equation}
    F=l L' + \theta J' -\frac{\epsilon(\theta)k_1}{\mu^{1/2}a^{-3/2}}\int_0^{l}r^2(l^*)\mbox{ d}l^*,
\end{equation}
where the primed quanities are the "new" (transformed) ones.\footnote{With the introduction of $\theta$ and $J$, the total energy of the system is formally conserved. Also, the "new" and "old" Hamiltonians are the same, since $\mathcal{H}'=\mathcal{H}+\frac{\partial F}{\partial t}=\mathcal{H}$.} The action $L$ up to $\mathcal{O}(\epsilon)$ accuracy reads
\begin{equation}
    L=\frac{\partial F}{\partial l}=L'-\frac{\epsilon(\theta)k_1}{\mu^{1/2}a^{-3/2}}r^2(l),
\end{equation}
from which
\begin{equation}\label{eq:solution_vonzeipel}
    a=\frac{1}{\mu}\left( L_1-\frac{\epsilon(\theta)k_1}{\mu^{1/2}}a^{1/2}\left( 1-e \cos E \right)^2 \right),
\end{equation}
where $r=a(1-e\cos E)$ is used, with $E$ being the eccentric anomaly. Note that $L'$ is constant up to the prescribed accuracy. 

\begin{figure*}
\centering
\includegraphics[scale=0.6]{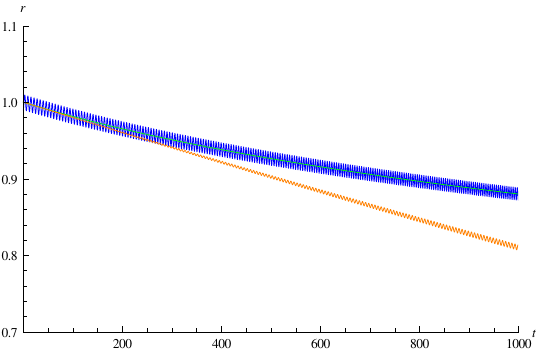}
\caption{Same as the left panel of Fig.~\ref{fig:harmonic}, but with $e_0=10^{-2}$. The orange curve is the solution of Eq.~\eqref{eq:solution_vonzeipel}, where we could use the approximation $E\approx l$, because $e_0\ll 1.$}
\label{fig:harmonic_vonzeipel}
\end{figure*}

This solution, plotted in orange in Fig.~\ref{fig:harmonic_vonzeipel}, is inconvenient for several reasons: (i) it is only first-order in $\epsilon$, although higher-order solutions can be obtained with more work via Deprit's method \citep{deprit}; (ii) it is not secular in the sense that it contains oscillatory terms of orbital period; (iii) the function $r(t)$ cannot be given explicitly in closed form.

\section{The smallness of the perturbation}\label{sec:app_smallness}

We take a Hamiltonian with a power-law perturbation,
\begin{equation}
    \mathcal{H}=\frac{\dot{x}^2+\dot{y}^2}{2}-\frac{\mu}{r}+\epsilon k r^n,
\end{equation}
where $k$ is a dimensional constant. The unperturbed (Keplerian) and perturbing energies, respectively, are

\begin{equation}
    E_0=-\frac{\mu}{2a}    
\end{equation}
and
\begin{equation}
    E_1\approx \epsilon k a^n,
\end{equation}
where orbit-averaging is executed and we use $\approx$ sign for an order-of-magnitude equality (dropping a factor of eccentricities, etc.). The effective mass parameter is
\begin{equation}
    \mu_\mathrm{eff}\approx \epsilon k a^{n+1}.
\end{equation}

Their time evolution, via Eq.~\eqref{eq:dot_a_short}, is driven by 
\begin{equation}\label{eq:e0dot_app}
    \dot{E_0}=\frac{\mu}{2a^2}\dot{a}\approx-\frac{\mu a \dot{\mu}_\mathrm{eff}}{2a^2\mu_\mathrm{eff}},
\end{equation}
\begin{equation}
    \dot{E_1}\approx\dot{\epsilon} k a^n
\end{equation}
and
\begin{equation}\label{eq:mudot_app}
    \dot{\mu}_\mathrm{eff}\approx \dot{\epsilon}a^{n+1}.    
\end{equation}
Note that we omit terms like $\epsilon\dot{a}\approx (\epsilon(0)+\dot{\epsilon}t)\dot{\epsilon}$, because they are of second order in small parameters like $\epsilon(0)$ (see Sec.~\ref{sec:discussions}) and $\dot{\epsilon}/\epsilon$ (see Sec.~\ref{sec:example_harmonic}). Substituting Eq.~\eqref{eq:mudot_app} back to Eq.~\eqref{eq:e0dot_app}, we see that $|\dot{E_0}|\approx|\dot{E_1}|\approx\dot{\epsilon}a^n$, i.e., they evolve with the same rate. Consequently, if their initial ratio was small, it remains so on timescales longer than $\epsilon(0)/\dot{\epsilon}$.

\bibliographystyle{abbrv}
\bibliography{cit}

\end{document}